\documentclass[acus]{JAC2000}

\usepackage{graphicx}

\begin{document}

\title{\flushright{THAP045}\\[15pt] \centering A COST-EFFICIENT PC BASED FRAME GRABBER AS BEAM DIAGNOSTIC TOOL IN AN EPICS ENVIRONMENT\thanks{Funded by the Bundesministerium f\"ur Bildung, Wissenschaft, Forschung und Technologie (BMBF) and the Land Berlin}}

\author{B. Franksen, R. Bakker, B. Kuner, J. Rahn, H. R\"udiger, BESSY, Berlin, Germany}

\maketitle

\begin{abstract}

Analysis of video images is a common way to determine the properties of an electron or synchrotron radiation beam. At BESSY a cost-efficient frame grabber hardware controlled by a Windows PC allows one to perform such an analysis in real-time. Live video images on the PC monitor give operators a direct view of the beam, optionally enhanced by artificial coloring and graphical overlays. The application software has been integrated into the EPICS \cite{Epics} environment by including a Channel Access (CA) server \cite{Ca}. Thus the numerical results and the application's control commands are accessible over a network through CA Process Variables. A newly developed C++ class library is used to provide the variables with the usual attributes required by advanced CA clients such as graphical display managers.

\end{abstract}

\section{INTRODUCTION}

In accelerator facilities the quality of the service depends among other factors on the geometrical properties of the beam, the most interesting aspect being its cross-section profile. It is thus common to install cameras and to display beam video images on some screens in the control room. Single images are often digitized and saved to disk for later processing by specialized image analysis tools. However, there are applications for which this is not sufficient, because the results of the image analysis need to be available in real-time.\footnote{Whenever used in this paper, the term `real-time' refers to `soft real-time', i.e. occasional violations of the timing constraints are acceptable as long as they are not exceeded on average.}

One of these applications is to enhance the optical qualities of the live video images as a human-readable diagnostic tool by artificial coloring and by adding graphical overlays. Others include feedback systems that automatically detect deviations from the desired geometry and correct them if possible or otherwise signal an alarm condition. Such systems are under consideration in response to the constantly rising demands for further improvement of the beam quality. In all cases a well-defined set of scalar variables that characterize the geometrical properties of the beam profile needs to be extracted from the huge amount of raw pixel data.

The standard setup for such a system incorporates a monochrome\footnote{For technical reasons, acquiring color images is normally not considered useful.} CCD camera (to produce the video data stream), a frame grabber card (to digitize the video data), and finally some software to perform the online image analysis and to enable interaction with users through high-level client applications.

This paper concentrates on two specific problem areas that appeared during the design and implementation of such a system at BESSY for which we found solutions that might be of general interest.

First, a suitable set of numerical values that characterize the beam profile geometry had to be found, with the constraint that these must be computable in real-time for high-resolution images arriving at a typical rate of ten frames per second. The application of statistical methods made it possible to achieve this goal. This will be discussed in Section \ref{Statistics}.

Second, the solution had to be integrated into an EPICS based control system, while remaining cost-efficient and adaptable to changing demands.\footnote{One the reasons being that the high-level applications still remain to be specified.} Since cost-efficiency rules out VME based hardware in favor of standard Windows PC technology, integration became a non-trivial issue. We discuss this in greater detail in Section \ref{Integration}.

\section{USING STATISTICAL METHODS FOR EFFICIENT IMAGE ANALYSIS}
\label{Statistics}

\subsection{Requirements}

Modern CCD cameras and frame grabbers deliver images with a resolution in the order of 400,000 pixels per frame. The typical rate at which video images are delivered and grabbed is around 10 Hz. Interesting geometrical properties of the beam include position, diameter (in various directions) and orientation (if the profile is not circular). These should be extracted from the raw pixel data without loss of the precision provided by the hardware, such that each frame can be processed and displayed on a computer monitor in 1/10 of a second. Processing includes artificial coloring as well as the addition of graphical overlays to indicate the results of the image analysis in a more intuitive way. We can expect the image to resemble a more or less elliptically shaped `smeared' spot, superimposed by varying amounts of background noise.

These demands immediately rule out a number of methods for image analysis (such as fitting or edge detection), simply because they cannot be implemented efficiently enough. Instead, the expected geometry suggests a much simpler approach using statistical methods, the application of which resulted in a system that is able to meet the required timing constraints on a 500 MHz Pentium PC under Windows NT 4.0.

\subsection{Statistical Image Analysis}

The main idea is to interpret the image as a two-dimensional probability distribution. With this identification, the center of the beam corresponds to the distribution's first moment, the mean value. In order to find out how far the beam is `smeared out' and in which direction, we also need to consider the second moments. For a two-dimensional distribution, these are given by the elements of the two-by-two covariance matrix, i.e.\ the four values $cov_{ij}$ with $i,j\in\{x,y\}$. The diagonal elements $var_i=cov_{ii}$ are the {\em variances}, the respective square roots $\sigma_i=\sqrt{var_i}$ the {\em standard deviations} in the direction of $i$, where $i$ may be either $x$ or $y$.

The single fact that the covariance matrix is {\em symmetric}, i.e. that $cov_{xy}=cov_{yx}$, implies that it can be diagonalized by singular value decomposition (SVD) and furthermore, that any pair of linearly independent eigenvectors are mutually orthogonal. There is exactly one angle in the interval $[0,\pi/2[$ such that after applying a rotation around this angle, the mixed covariances become zero and the two variances become extreme (with respect to the angle of rotation), i.e. one will be minimal, the other one maximal. This shows that any two-dimensional distribution (assuming it is not completely direction independent) has exactly one direction of maximal deviation and one of minimal deviation, and that these directions are mutually orthogonal and can be determined by an SVD of the covariance matrix. The latter has to be computed only once using any arbitrarily chosen coordinate system.

The crucial point here is that the mere computation of five scalar values -- two for the mean value, three for the covariance matrix -- implicitly contains information about the deviations in {\em all} directions, i.e. the exact shape of the so called {\em ellipse of standard deviation}. If the distribution of the beam intensity is nearly gaussian (as we can normally expect it to be), this ellipse is a good approximation of the geometry of the beam profile. Moreover, these five numerical values can be computed very efficiently.

\subsection{Efficient Calculation}

The reason for this is twofold: First, they can be accumulated in one single pass over the image. Second, during that single pass, only simple arithmetical operations need to be performed. More specifically, for each frame it suffices to accumulate the six values $p$, $px$, $py$, $px^2$, $py^2$, and $pxy$ for each pixel value $p$ at coordinates $x$ and $y$. With standard camera resolutions, a monochrome color depth of 8 bits per pixel, and a compiler which supports 64 bit integer emulation (like Microsoft's Visual C++ does), floating point operations can be completely avoided inside this loop.

The additional computations necessary e.g. for calibration and SVD are much less critical, since they need to be performed only once per frame -- instead of once per pixel. Analytical solution of the SVD gives the eigenvalues of the covariance matrix (or equivalently the maximum and minimum deviations) as
\[
    \frac{var_x+var_y}{2} \pm \sqrt{
        \left(\frac{var_x-var_y}{2}\right)^2+cov_{xy}
    },
\]
which are assumed at the angles
\[
    \frac{1}{2} \arctan\left(\frac{2\,cov_{xy}}{var_x-var_y}\right)+
        n\,\frac{\pi}{2}\quad (n=0,1,2,\dots).
\]

\subsection{Background Noise}

An important factor when using statistical methods as described above is the amount of background noise. Even if the beam's contribution to the signal is near to the ideal gaussian distribution, additional noise will make the deviation from the center appear larger than it actually is. This effect must be compensated for, especially since the level of noise has the tendency to depend on various external conditions, including the ring current, i.e. beam intensity. Three methods are presented here that can be used to deal with this problem. The system developed at BESSY provides all three methods, configurable by the user or high-level applications.

One rather simple approach is to cut off any signal below a certain threshold, the height of which must be determined experimentally. While being very efficient to implement, this approach is of limited use if the noise level varies too much.

A more advanced approach is to dynamically determine the level of background by an integration over a suitable closed path around the beam, typically the border of the image or the so called {\em area of interest} to be analyzed. This value will be subtracted from the whole area. This approach works well if the noise is evenly distributed over the area. It is not too costly to implement, since the number of pixels to be scanned is far lower than that of the whole area.

A third method might be appropriate in situations where strong background radiation is present even without a beam. In this case, it makes sense to record an image of this background and to subtract it pixel by pixel from the image to be analyzed. This approach implies a greater penalty on efficiency and rules out automatic gain control by the camera. Nevertheless, there appear to be situations where this is the only way to get reliable results.

\section{INTEGRATION INTO AN EPICS ENVIRONMENT}
\label{Integration}

EPICS uses a communication protocol called Channel Access (CA), which is specialized for efficient communication in large distributed control systems. It follows the client-server paradigm in that low-level I/O controllers (IOCs) are servers for process variables (PVs), while high-level applications (e.g. user interfaces) act as clients that monitor, query, or update these variables.\footnote{This is a somewhat simplified description. For details, see \cite{Epics}.} Normally, the server side is implicitly programmed by the configuration of a runtime database.

Thus, implementing the image analysis tool on an IOC would have been the easiest way to get CA server funcionality. Since the current EPICS release only supports VME/VxWorks as target platform for IOCs, the solution would have to be based on a VME frame grabber card. Unfortunately, the market for these cards is very thin and the available hardware extremely expensive. Cost-efficient frame grabbers are found almost exclusively in the PC market and drivers are normally only available for MS Windows.

Since, for obvious reasons of reliability and maintenance cost, extending our infrastructure to support Windows PCs was out of the question, it was decided to make the PC/Windows solution a stand-alone system. Communication with the users through high-level applications was restricted to a well-defined set of CA variables. This was possible due to some recent development efforts by the EPICS community regarding the so called {\em Portable CA Server (CAS)} library.

\subsection{The Portable CA Server Library}

This library, written in C++, enables developers to program a CA server on any of the supported host platforms (including MS Windows) independently of the runtime database support provided by the EPICS IOC core. However, due to some historical ballast,\footnote{This refers mainly to the so-called Generic Data Descriptor (GDD, see \cite{Ca}) a C++ development originating from the time when the first C++ compilers appeared that did not even support virtual functions.}, using the CAS library is not as straight-forward as the apparently nice and clean interface seems to suggest. This is especially true if the server tool is supposed to deliver not only the plain values of its PVs, but also the usual {\em attributes} associated with these values, such as alarm and display limits, timestamps, alarm status and alarm severity. Support for attributes is essential because they are routinely requested by high-level applications such as display managers. In contrast to the traditional IOC case, where all this functionality is already taken care of by the runtime system, providing these attributes through the portable CAS is the tedious and error-prone task of the programmer.

It was decided that a clean solution to this problem would involve some effort and that to make this effort more worthwhile, it should be generic enough to be usable as an independent library on top of the CAS. Starting from an existing code base written by Kay-Uwe Kasemir from LANL, the final result was a class library named XCas \cite{XCas}.

\subsection{The XCas Library}

Being easy to use, not requiring expert knowledge of C++, was one of the major design goals. This necessarily included shielding the user from any low-level details of the CAS interface. PVs are automatically provided with all attributes suitable for the chosen data type. Adding a process variable to a program requires nothing more elaborate than the declaration of an object of one of the supplied classes. In fact, the classes were modelled after the standard EPICS database records; consequently the base class was named {\em Record} and the standard PV attributes were implemented as record {\em Field}s, including the aspect of being not only accessible as attributes to some PV but also as independent PVs in themselves. This enables clients to directly control or monitor these attributes in a way similar to runtime database records and their associated fields. The classes provide complete functionality and are not normally meant to be subclassed (although they can be, if desired). Facilities to install callbacks to react to external updates of PVs are implemeted using class templates to avoid the usual typecasts. The server itself acts completely hidden and is set up and invoked only by internal methods of class {\em Record} and its (few) specialized descendants.

\section{CONCLUSIONS}

We have shown that by using statistical methods the image analysis of high-resolution video data can be implemented efficiently enough to meet the real-time requirements arising from the adoption of advanced feedback control and error detection facilities that rely not only on the beam position but also on parameters to identify the shape and direction of the beam's cross-section profile.

We have further shown that an image analysis tool capable to meet these requirements can be based on a target platform with a competitive market. The Portable CA Server library can be effectively used for the integration into an EPICS control system without exposing the latter to the shortcomings in reliability and maintainability commonly associated with the chosen platform. The effort to decouple such a tool from the intricacies of dealing with the more difficult aspects of the interface to the Portable CAS has resulted in an independent class library, which can be used to simplify the process of writing any sort of CA server tools.

\end{document}